\documentclass[journal]{IEEEtran}
\usepackage[latin1]{inputenc}
\usepackage[T1]{fontenc}
\usepackage{verbatim}
\usepackage{amsfonts,amsmath,epsfig,bbm,subfigure,graphics,cite,psfrag,url,array,subfigure}

\newtheorem{proposition}{Proposition}
\newtheorem{assumption}{Assumption}

\def\N{{\mathbb N}}

\def\P{{\mathbb P}}
\def\E{{\mathbb E}}

\def\P{{\mathbb P}}

\def\cal{\mathcal}
\def\k{k}
\def\Q{{\mathbb Q}}
\def\p{p}

\begin{document}
%
\title{Inference of Flow Statistics via Packet Sampling in the Internet}
\author{
Yousra Chabchoub, Christine Fricker, Fabrice Guillemin, and Philippe Robert
\thanks{Y. Chabchoub, C. Fricker and P. Robert are with INRIA, Domaine de Voluceau, 78153 Le Chesnay, France,
Email:\{Yousra.Chabchoub, Christine.Fricker,Philippe.Robert@inria.fr\}} 
\thanks{F. Guillemin is with Orange Labs, 2 Avenue Pierre Marzin, 22300 Lannion, France, Email:Fabrice.Guillemin@orange-ftgroup.com}
}

\maketitle

\begin{abstract}
We show in this note that by deterministic packet sampling, the  tail of the distribution of the original flow size  can be obtained by rescaling that of the  sampled flow size. To recover information on the flow size distribution lost  through packet sampling, we propose some heuristics based on measurements from different backbone IP networks. These heuristic arguments allow us to recover the complete flow size distribution.
\end{abstract}

\begin{keywords}
Packet sampling, Flow statistics, Pareto distribution.
\end{keywords}

\IEEEpeerreviewmaketitle

\section{Introduction}

Packet sampling is an efficient method of reducing the amount of data to analyze when performing measurements in the Internet. The simplest and the most popular packet sampling technique consists of selecting one  packet every other $k$ packets. This technique is referred to as deterministic 1-out-of-$k$ sampling in the technical literature and has notably been implemented  in CISCO routers \cite{Netflow}. Even if this sampling scheme suffers from several drawbacks, identified for instance in \cite{estan}, it is widely used in today's operational networks. 

The basic problem of packet sampling is that it is difficult to infer the original flow statistics from sampled data. Note that a flow is defined as the set of those packets sharing some common addressing information, typically the same source and destination IP addresses, the same source and destination port numbers together with the same protocol type. 

Flow statistics inference from sampled data has been addressed in previous studies. Duffield \emph{et al} \cite{Duffield,Duffield2} study the accuracy of different estimators based on multiplying  the sampled flow size by the sampling factor $k$, but their method does not apply to the complete range of the flow size. Hohn and Veitch \cite{Hohn} use generating function techniques to invert the flow size distribution but the proposed procedure is numerically unstable. Mori \emph{et al} \cite{Mori} use a Bayesian approach to inferring the characteristics of long flows.

In this paper, we develop a probabilistic approach to inverting sampled traffic together with some heuristic arguments.  First, we note that when  observing sampled traffic, we can only compute the distribution of the random variable $\tilde{v}$ describing the number of packets in sampled flows and $K_s$ the number of sampled flows. If there are originally $K$ flows, we have
\begin{equation}
\label{deftildev}
\P(\tilde{v} = j) = \frac{1}{K_s}\sum_{i=1}^K I_{i,j} 
\end{equation}
where $I_{i,j} = 1$ if the $i$th flow has been sampled $j$ times and $I_{i,j}=0$ otherwise.

Under some reasonable assumptions on the sampling process, we show in this note that the tail of the original flow size distribution can be obtained by rescaling the distribution of  the sampled flow size distribution. It is however not possible to totally recover the original flow size distribution because information on small or moderate flow sizes is loss through sampling. To overcome this problem, we propose some heuristic arguments based on measurements and exploiting  a priori information on flows.  We consider here TCP traffic only.

The rest of this note is organized as follows: In Section~\ref{assumptions}, we make some reasonable assumptions on the sampling process. In Section~\ref{proof}, we prove that the tail of the original flow size can be obtained by rescaling that of the sampled flow size. In Section~\ref{heuristics}, we present some heuristic arguments to recover the total flow size distribution. Concluding remarks are presented in Section~\ref{conclusion}.

\section{Assumptions on the sampling process}
\label{assumptions}

When observing in a time window of length $\Delta$ traffic on a high speed link,  one may
reasonably assume that the packets of the different active flows are sufficiently
interleaved. Hence, one may suppose the selection of packets among active flows at a
sampling time is random. 

Moreover, in a time window of length $\Delta$, flows start and finish and some of them may
be silent (for instance in the case of flows alternating between On and Off periods). In
\cite[Section~3.3]{Chabchoub:01}, it is shown that these fluctuations may neglected at the
first order (i.e., when computing mean values) and it can be assumed that flows are
permanent. Under the two above assumptions, we suppose that the probability of selecting a
packet of a given flow, say, flow $i$, is  equal to $v_i/V_i$, where $v_i$ is the size of
flow $i$ and $V_i$ is the total number of packets arrived when flow $i$ is active. 

\section{Tail of the sampled  flow size}
\label{proof}

Let $W_j \stackrel{\mathrm{def}}{=}\sum_{i=1}^K I_{i,j}$, the number of flows sampled $j$ times.
\begin{proposition}\label{boundobs}
If $K$ flows are active during a time window of length $\Delta$, the mean value $\E(W_j)$  satisfies
\begin{align}\label{asymp}
\left|{\E(W_j)} - K \Q_{j}\right|\leq \p \sum_{i=1}^K\E\left({v_i^2}/{V_i} \right),
\end{align}
where $\p=1/\k$, $v_i$ is the random number of packets in a flow, and $\Q$  is the probability distribution defined by 
\begin{align}
\label{defQ}
\P(\Q = j) \stackrel{def}{=} \Q_{j}=\E\left(\frac{{\left(\p v\right)}^j}{j!}e^{-{\p}v}\right),
\end{align}
\end{proposition}

\begin{proof}
Let us condition  on the values of the set ${\cal F}=\{v_1, \ldots, v_K, V_1, \ldots,V_K \}$. Under the  assumptions of Section~\ref{assumptions}, the number of times that the  $i$th flow is sampled  is  equal to the sum
\[
S_i=  B_{1}^i+B_{2}^i+\cdots+B_{\p V_i}^i,
\]
where $B_\ell^i$ is equal to one if the  $\ell$th sampled packet is from the $i$th flow, which event occurs with probability $v_i/V_i$. The random variables  $(B_\ell^i, \ell \geq 1)$ are i.i.d. Bernoulli random variables and Le Cam's Inequality \cite{Barbour}  then states
\[
\left\|\P(S_i    \in \cdot )-\P \left(Q_{\E(S_i)} \in \cdot \right)\right\|_{tv} \leq \sum_{\ell=1}^{pV_i} \P(B^i_\ell =1)^2,
\]
where $\|.\|_{tv}$ is the total variation norm and $Q_{\E(S_i)}$ is a Poisson random variable with mean $\E(S_i)$. By deconditioning  with respect to the set $\mathcal{F}$,   we have by using the distribution $\Q$
\begin{equation}
\label{distWj}
\left\|\P(S_i\in\cdot)-\Q\right\|_{tv} \leq \p\E\left({v_i^2}/{V_i}\right).
\end{equation}
In particular, for $j\in\N$, $\left|\P(S_i=j)-\Q_j\right| \leq \p\E\left({v_i^2}/{V_i}\right)$. Since $\E(W_j)=\sum_{i=1}^K\P(\tilde{v}_i=j)$,
summing on $i$ yields Equation~\eqref{asymp}.
\end{proof}

If $K$ is sufficiently large, we have from Equation~\eqref{deftildev} and the above proposition, we have for $j \geq 1$
\begin{equation}
\label{sampleprob}
\P(\tilde{v}=j) \sim \frac{1}{\nu}\frac{\E(W_j)}{K},
\end{equation}
 where $\nu =K_s/K$ is the probability of sampling a flow.

\begin{proposition}
\label{asymptail}
If all flows have a negligible contribution to the total volume of traffic (i.e., $\E(v_i^2/V_i)\ll 1$ for all $i=1,\ldots,K$), if $K$ is sufficiently large,  and if the flow size distribution has a slowly varying tail, then when $j \to \infty$
\begin{equation}
\label{estimvhat}
\P(\tilde{v}\geq j) \sim \P\left(v\geq \frac{j}{p}\right)/\nu.
\end{equation}
\end{proposition}

\begin{proof}
From Equation~\eqref{sampleprob}
\begin{equation}
\label{approxprob}
\P(\tilde{v}=j) \sim \frac{1}{\nu}\E\left( \frac{(pv)^j}{j!} e^{-pv} \right)  = \frac{p^j}{j! \nu} \sum_{\ell=0}^\infty e^{j\log\ell -p\ell}\P(v=\ell) .
\end{equation}
Consider the sum  $\sum_{\ell = 0}^\infty e^{f(p,\ell)}\P(v=\ell)$, where $f(p,x) = j\log x  -p x$, which is maximum at point $j/p$. By assuming that  the function $\ell \to \P(v=\ell)$ is heavy tailed, Laplace method gives for large $j$ 
$$
\E\left( \frac{(pv)^j}{j!} e^{-pv} \right) \sim \frac{p^j}{j!} e^{f(p,j/p)}\P\left(v=\frac{j}{p}\right)\sum_{\ell =-\infty}^\infty e^{\ell^2\frac{f''(p,j/p)}{2}},
$$
where $f''(p,j/p) = -p^2/j$ and $e^{f(p,j/p)} = (j/p)^je^{-j}$. If $j$ is sufficiently large, Stirling formula gives $j! \sim \sqrt{2\pi}j^{j+\frac{1}{2}}e^{-j}$. In addition, from \cite{Abramowitz}, we have $\sum_{n=-\infty}^\infty e^{-a n^2} \sim \sqrt{{\pi}/{a}}$ when $a \to 0$. This implies that $\sum_{\ell =-\infty}^\infty e^{\ell^2\frac{f''(p,j/p)}{2}} \sim \frac{\sqrt{2\pi j}}{p}$ and then $\P(\tilde{v}=j) \sim {\P(v=j/p)}/{p}$, when $j$ is sufficiently large. Since
$$
\P(\tilde{v}\geq j) \sim \frac{1}{\nu p}  \sum_{k=j}^\infty \P(v =k/p)   \sim \frac{1}{\nu} \int_j^\infty d\P(v=k/p),
$$
Equation~\eqref{estimvhat} follows. \end{proof}


\section{Heuristics for the total flow size distribution}
\label{heuristics}

Proposition~\ref{asymptail} shows that the tail of the  complementary cumulative distribution function (ccdf) of the original  flow size can be obtained by rescaling that of the sampled flow size. We can however verify through examples that information on that  distribution for small or moderate flow size values is lost. 

\begin{figure}[hbtp]
\begin{center}
\subfigure[Original size.\label{FTflowsize}]{\scalebox{.45}{\input{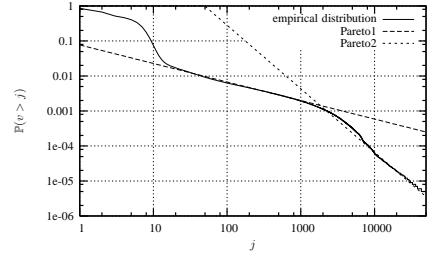}}}

\subfigure[Sampled size ($p=1/100$).\label{sampledFTflowsize}]{\scalebox{.45}{\input{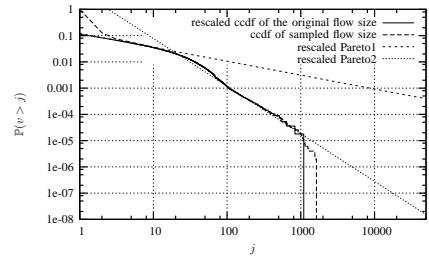}}}
\end{center}
\caption{Flow size distribution in the France Telecom ADSL trace.}
\end{figure}

We exemplify this phenomenon by considering a 2 hour long real traffic trace from a 1~Gbit/s transmission link of the France Telecom IP backbone network carrying ADSL traffic. The original flow size is depicted in  Figure~\ref{FTflowsize}  and the deterministically sampled flow size in Figure~\ref{sampledFTflowsize}, which exhibits good agreement with the rescaled distribution $\P(v=j/p)/\nu$ for sufficiently large $j$ as predicted by Proposition~\ref{asymptail}.  But all  information for moderate values of the flow size  is contained in a few values, in this case  $\P(\tilde{v}\geq j)$ for $j=2,3$. The same phenomenon (see \cite{Chabchoub:02}) has been observed for an Abilene traffic trace available at http://pma.nlanr.net/Traces/Traces/long/ipls/3/.

In fact, through numerous experiments with real traffic traces, it has been observed in \cite{Chabchoub:02} that $\P(v \geq j/p)$ can be approximated by $\nu \P(\tilde{v}\geq j)$ when  $j\geq j_0$ for some $j_0>0$. The problem is then to estimate the quantities $\P(v=j)$ for $j =1,\ldots,j_0/p-1$. 

We have from Equation~\eqref{approxprob}
\begin{equation}
\label{linsys}
\P(\tilde{v}=j) \sim \frac{1}{\nu}\sum_{\ell=1}^\infty \frac{(p\ell)^j}{j!} e^{-p\ell} \P(v = \ell) 
\end{equation}
and we know by Proposition~\ref{asymptail} that for $j\geq j_0$, this equation is equivalent to $\P(\tilde{v}=j_0) = \P(v=j/p)/(\nu p)$. It follows that for determining the ($j_0/p-1$) quantities $\P(v=\ell)$ for $\ell = 1, \ldots,j_0/p-1$, we have only $j_0$ equations. The problem is hence clearly under-determined. Some heuristics are needed to recover the complete flow size distribution.

It has been observed in \cite{Chabchoub:02}  that depending on the size of the observation window $\Delta$, the sampled flow size distribution can locally be approximated by means of Pareto distributions. This leads us to  make the following assumption.
\begin{assumption}
\label{assump1}
There exist some $m>0$ and some integers $j_0 < j_1 <  ... < j_m=\infty$  such that for $\ell=1,\ldots,m$ and $j \in [j_{\ell-1},j_\ell]$, $\tilde{v}$ has a Pareto distribution of the form
$$
\P(\tilde{v}\geq j) = \P(\tilde{v}\geq j_{\ell-1}) \left({j_{\ell-1}}/{j}\right)^{a_\ell}
$$
for some shape parameter $a_\ell >0$.
\end{assumption}

When $\Delta$ is adequately chosen, the tail may be  uni-modular (i.e., $m=1$), but when $\Delta$ is too large, we can have $m >1$. For the above France Telecom trace ($\Delta = 2$ hours), $m=2$ as shown in Figure~\ref{sampledFTflowsize}. 

By using Proposition~\ref{asymptail}, we deduce that  for $\frac{j_{m-1}}{p}\leq j \leq \frac{j_{m}}{p}$
\begin{equation}
\label{form1}
\P(v \geq j) \sim {\nu}\P\left(\tilde{v}\geq j_{m-1}\right) \left({j_{m-1}}/(pj)\right)^{a_m},
\end{equation}
The above equation implies that $\P(v \geq j) $ can locally be approximated by a Pareto distribution with shape parameter $a_m$, as shown in Figure~\ref{FTflowsize}.

For inferring the quantities $\P(v=j)$ for  $j=1, \ldots, j_0/p-1$, we need more
assumptions. Numerous experiments \cite{Chabchoub:02} have shown that when  $j < b_0$ for
some $b_0>0$, $\P(v = j)$ follows a geometric distribution. 

\begin{assumption}
\label{assump2}
There exists some $b_0>0$ such that for $1\leq j <b_0$, $\P(v = j)=(1-r)r^{j}$ for some $r>0$.
\end{assumption}

\begin{figure}[hbtp]
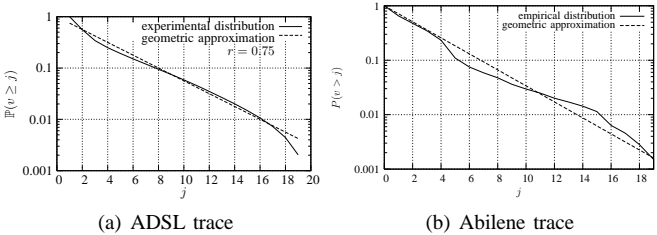

\begin{center}
\subfigure[ADSL trace\label{miceFT}]{\scalebox{.33}{\input{micepkt.pslatex}}}
\subfigure[Abilene trace\label{miceAbilene}]{\scalebox{.35}{\large \input{miceAbileneB.pslatex}}}
\end{center}
\caption{Ccdf of the number of packets in flows with less than $b_0=20$ packets.}
\end{figure}

The above assumption is supported by experiments, as shown in Figure~\ref{miceFT} and \ref{miceAbilene} for the France Telecom and Abilene traffic traces, respectively. The value $b_0=20$ has been successfully tested in numerous experiments. 

By using Equation~\eqref{form1} and Assumption~\ref{assump2}, we have the form of the distribution for $j \leq b_0$ and $j \geq j_0/p$. To fill the gap, we use the following heuristic: $\P(v\geq j)$ for $b_0\leq j\leq j_1/p$ has the same form as in Equation~\eqref{form1}, namely $\P(v\geq j)= \P(v\geq b_0)(b_0/j)^{a_1}$. Equation~\eqref{linsys} can then be rewritten as
\begin{multline}
\label{form2}
\P(\tilde{v} = j)  \sim \frac{\P(v<b_0)}{\nu}\sum_{\ell=1}^{\infty}(1-r)r^\ell  \frac{(p\ell)^j}{j!} e^{-p\ell}    \\ +  \frac{1}{\nu}\sum_{\ell=b_0}^{\infty} \frac{(p\ell)^j}{j!} e^{-p\ell} \P(v = \ell).
\end{multline}

The shape parameters $a_\ell$ for $1\leq \ell \leq m$ are determined from the sampled flow size distribution by using standard Maximum Likelihood Expectation (MLE) procedures. The parameter $b_0$ is set equal to 20; this choice is purely phenomenological but corresponds to the number of packets needed to leave the slow start regime with a maximum window size of 32 Kbytes. The parameter $\P(v\geq b_0)/\nu$ is obtained by using Proposition~\ref{asymptail}, namely by computing the ratio $\eta\stackrel{def}{=} \P(\tilde{v}\geq j)/(b_0p/j)^{a_1}$  for $j \in \{j_0,\ldots, j_1\}$, which is by assumption independent of $j$. The number of flows with at least  $b_0$ packets is $K_0^+=\eta K_s$. Equation~\eqref{form2} multiplied by $K_s$ for $j=1,2$  is then used to compute the parameter $r$ and the number $K_0^-$ of flows with less than $b_0$ packets. 
The total number of flows is then $K=K_0^++K_0^- $ and the probability of sampling a flow is estimated by the ratio  $K_s/K$.

By using the above method for the France Telecom ADSL trace with $p=1/100$, we find $j_0=3$ and the estimated shape parameters $\hat{a}_1= .54$ and $\hat{a}_2=1.81$, which are close to the experimental values $a_1 = .52$ and $a_2 =1.81$ for the original flow size. We then find $\P(v\geq b_0)/\nu=.3$ and since  $K_s= 1,120,546$, we  obtain the estimate $\hat{K}_0^+=336,163$, while the actual value is $K^+_0=343,004$. By neglecting the term due to flows with at least 20 packets in Equation~\eqref{form2}, we then find the estimate $\hat{r}= 0.84$ while the actual experimental value is $r=.75$. This yields a number of flows with less than $b_0$ packets $\hat{K}_0^-\sim 20.1e6$ while the actual value is $K_0^-\approx 19.8e6$. Finally, the estimated total number of flows is $\hat{K}= 20.4e6 $ while the actual value is $K= 20.1e6$ and  we find the estimate $\hat{\nu}= .054$ for the probability of sampling a flow while the experimental value is $\nu = 0.057$.

\section{Conclusion}
\label{conclusion}

We have shown in this paper by using  probabilistic arguments that the original size distribution of large flows can be recovered from that of the sampled flow size. A critical parameter is nevertheless the flow sampling probability, which can be estimated only when the size of small flows is known. To overcome this problem, we argue that it is necessary to exploit a priori information on flows. By using this principle, we have shown that it is possible to recover the complete flow size distribution together with  the number of flows.


\end{document}